\documentclass[draftclsnofoot,onecolumn,12pt]{IEEEtran}
\usepackage{subfigure}
\usepackage{booktabs}
\usepackage{bm}
\usepackage{multirow}
\usepackage{array}
\usepackage{amsfonts}
\usepackage{float}

\usepackage[ruled,linesnumbered]{algorithm2e}

\usepackage{cite}

\ifCLASSINFOpdf
 \usepackage[pdftex]{graphicx}

\usepackage{amsmath}

\usepackage{algorithmic}

\usepackage{array}

\begin{document}

%
\title{Correlation-based Initialization Algorithm for Tensor-based HSI Compression Methods}
%
%
%

\author{Rui~Li,
        Zhibin~Pan and Yang~Wang
        \thanks{This work is supported in part by the Open Research Fund of Key Laboratory of Spectral Imaging Technology, Chinese Academy of Sciences (Grant No. LSIT201606D) and the Open Project Program of the National Laboratory of Pattern Recognition (NLPR) (Grant No. 201800030).}
\thanks{Rui~Li is with the School of Electronic and Information Engineering, Xi'an Jiaotong University, Xi'an, P. R. China, 
710049.}
\thanks{Zhibin~Pan is with the School of Electronic and Information Engineering, Xi'an Jiaotong University, Xi'an, P. R. China, 710049,
 (e-mail: zbpan@mail.xjtu.edu.cn).}
\thanks{Yang~Wang is with the School of Electronic and Information Engineering, Xi'an Jiaotong University, Xi'an, P. R. China, 710049.}
}

%
%

\markboth{under reviewing of Multimedia tools \& applications}%
{Shell \MakeLowercase{\textit{et al.}}: Bare Demo of IEEEtran.cls for IEEE Journals}
%



\maketitle

\begin{abstract}
Tensor decomposition (TD) is widely used in hyperspectral image (HSI) compression. The initialization of factor matrix in tensor decomposition can determine the HSI compression performance. It is worth noting that HSI is highly correlated in bands. However, this phenomenon is ignored by the previous TD method. Aiming at improving the HSI compression performance, we propose a method called correlation-based TD initialization algorithm. As HSI is well approximated by means of a reference band. In accordance with the SVD result of the reference band, the initialized factor matrices of TD are produced. We compare our methods with random and SVD-based initialization methods. The experimental results reveal that our correlation-based TD initialization method is capable of significantly reducing the computational cost of TD while keeping the initialization quality and compression performance.
\end{abstract}

\begin{IEEEkeywords}
Tensor decomposition (TD); hyperspectral image compression; initialization.
\end{IEEEkeywords}

%
\IEEEpeerreviewmaketitle

\section{Introduction}
With the wide applications of hyperspectral imaging technology, hyperspectral image (HSI) appears in many fields such as: remote sensing \cite{fang2017hyperspectral}, diagnostic medicine \cite{akbari2010detection}, image classification \cite{zhang2016weighted}, unmixing \cite{mei2018robust}, and more \cite{coutinho2017low,
shi20173d,md2017hyperspectral,
ma2018hyperspectral}. The HSI is composed by plenty of bands which are images of the same scene on different wavebands, Therefore, the bands of HSIs share the similar characteristics. However, hyperspectral imaging suffers from many issues. One major issue is that the large amount of data makes HSI difficult to transmit and process. In addition, noise and limited illumination in individual band are two problems. Aiming at addressing the mentioned issues, Tucker Decomposition (TD) is presented \cite{tucker1966some,
karami2012compression,
wang2017compressive,
sidiropoulos2017tensor,
zeng2017method}. The HSI is considered as a tensor while the decomposition results are factor matrices and a core tensor. The sizes of core tensor in TD should be smaller than original corresponding sizes. Accordingly, the number of eigenvectors is reduced. HSIs are significantly compressed. 

As traditional method tackles the spatial domain, the strong correlation between bands is ignored \cite{dragotti2000compression}, consequently, the compression performance of these methods is compromised for this drawback. One advantage of TD is that it learns a low dimensional representation. The spatial and spectral correlations are simultaneously addressed. The vector in factor matrices depicts the low dimension embedding of a corresponding slice of tensor \cite{jiang2018image}. Thus, the decomposition solves the spectral correlation in HSI well. However, there are some issues needing to be settled. The TD is solved by alternating least square (ALS) method. Although the alternating minimization in TD is nonconvex, the objective function is strictly convex with respect to one set of variables \cite{cichocki2008nonnegative}. The ALS method is influenced by many issues \cite{da2017initialization}, one critical issue is that the initial factor matrices are required. The convergence rate of TD is remarkably influenced by the qualities of initial factor matrices. A well designed initial factor matrix can make sure the convergence of iterative strategy and reduce the number of iterations. Otherwise, the algorithm may get stuck in a local minimum, for some data the convergence could be very slow when factor matrices are ill-conditioned or when collinearity occurs in the columns of these matrices \cite{cichocki2008nonnegative}.

There are two kinds of initialization method for TD. The first method is random method, which generates orthogonal vectors of factor matrix randomly. The factor matrix is columnwise orthonormal. The random method is easy to implement, but the TD may converge in many iterations, and the convergence of the TD is not guaranteed. The second method resorts singular vector decomposition (SVD) to tackle the factor matrices. The factor matrices associated to each mode are obtained by applying SVD on the corresponding matricization of the tensor. 
\begin{figure}
\centering
\subfigure[The averaged correlation of all bands in Indian pines.]{%
\resizebox*{7cm}{!}{\includegraphics{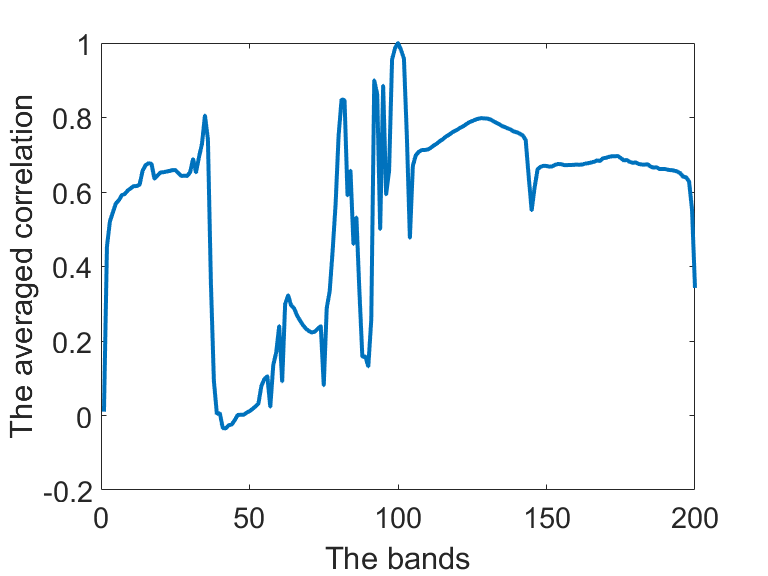}}}\hspace{5pt}
\subfigure[The averaged correlation of all columns in the 50th band of Indian pines.]{%
\resizebox*{7cm}{!}{\includegraphics{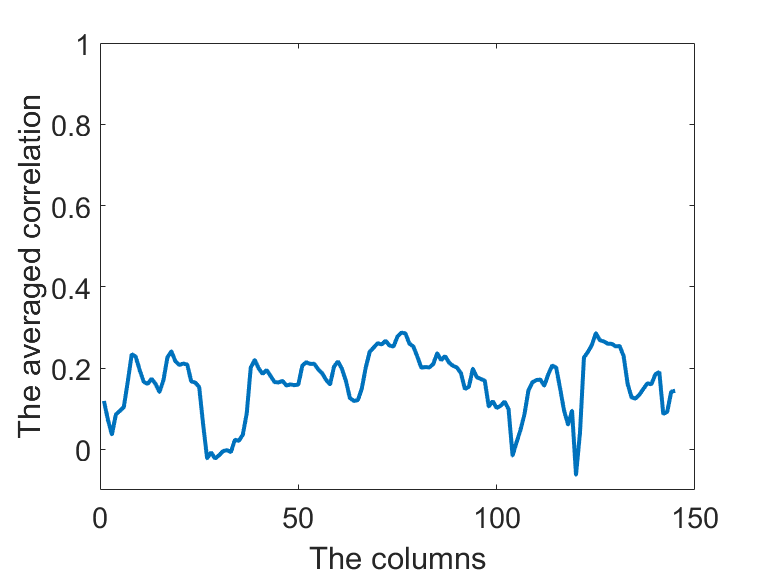}}}
\caption{The comparison of spectral correlation and spatial correlation in Indian pines.} \label{sample-figure}
\end{figure}

However, these two initialization methods are neither effective nor addressed for HSIs. Fig.1 shows that HSI has stronger spectral correlation, which implies the band in HSIs can be well represented by other bands. To the best of our knowledge, there is no study in literature utilize the spectral correlation in HSI. We can utilize the spectral correlation to design a new TD initialization method for HSI to improve the compression performance and reduce the time consumption.

The highlights of this paper are listed below:
\begin{itemize}
   \item We first explain the factor matrices in TD can be well approximated by the SVD results of some reference band;
  \item According to our conclusion we give our correlation-based initialization method for TD;
  \item We test our method to compress the HSI and achieve considerable results.
\end{itemize}

The rest of this paper are organized as follows: The basic knowledge of tensor analysis is introduced in Section 2. We prove that the factor matrices in TD can be well approximated by the SVD results of our reference band, and then propose our initialization method in Section 3. The experimental results are given in Section 4. We finally conclude our method in Section 5.

\section{Tensor Preliminary}
Before we introduce our method, we give some preliminaries of tensor analysis to help understand our method in this section.
The notations used in this paper are summarized in the table below.
\begin{table}[h]
   \centering
   \caption{Notations About Tensor}
   \label{tab1}
   \begin{tabular}{p{0.3\columnwidth}<{\centering}p{0.55\columnwidth}<{\centering}}
   \toprule
\multicolumn{2}{c}{Notations used in this paper}\\  
   \midrule
    $\otimes$  & Kronecker product \\
    $\bm{\mathcal{X}}\in\mathbb{R}^{{I_1} \times {I_2} \times \dots\times {I_N}}$ & A\ $N_{th}$-order tensor \\
    $x( {{i_1},{i_2},...,{i_N}} )$& The entry of tensor   $\bm{\mathcal{X}}\in\mathbb{R}^{{I_1} \times {I_2} \times \dots\times {I_N}}$ \\
    $\textbf{X}_{(1)}$ & The mode-1 matricization of $\bm{\mathcal{X}}$\\
    $\textbf{X}^{I\times{JK}}$&A matrix of size $I\times{JK}$\\
    $\textbf{U}^{(i)}$ & $i_{th}$-order\\
    $\textbf{X}_{(i)}^k$ & The $k_{th}$ slice along mode-$i$ in tensor $\bm{\mathcal{X}}$\\
    $\textbf{X}_{::i}$ &The $i_{th}$ slice in tensor $\bm{\mathcal{X}}$\\
    $\Vert\bm{\mathcal{X}}\Vert_{F}$ &The Frobenius norm of a tensor $\bm{\mathcal{X}}$\\
 \bottomrule
   \end{tabular}
\end{table}

In multilinear algebra, Tensor is actually a multidimensional array. We denote $N_{th}$-order tensors by bold uppercase mathcal letters $\bm{\mathcal{X}}\in\mathbb{R}^{{I_1} \times {I_2} \dots\times {I_N}}$.  The order of the tensor is also referred as mode. The first-order tensor is known as vector and the second-order tensor is known as matrix. The mode-n matricization means rearranging the tensor as matrix by fixing the $n_{th}$-order index and varying other indices. It is also a flattening process in Fig. 2. For a third-order tensor $\bm{\mathcal{X}}\in\mathbb{R}^{{I} \times {J} \times {K}}$, there are three mode-n matricizations.

The mode-1 Kiers matricization \cite{kolda2009tensor} of a third-order tensor $\bm{\mathcal{X}}$  results in $\textbf{X}_{(1)}$. $\textbf{X}_{(2)}$ and $\textbf{X}_{(3)}$ are the mode-2 and mode-3 Kiers matricization results, respectively. The matricizations in this paper are all Kiers matricizations.
\begin{figure}[tpb]
\centering
\includegraphics[width=0.8\textwidth]{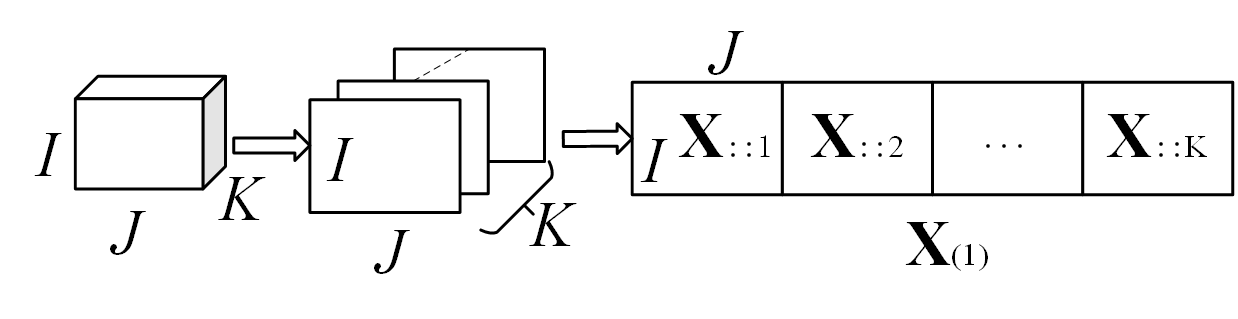}
\caption{The mode-1 matricization is obtained by flattening the tensor along its first mode.}
\label{fig1}
\end{figure}

The Frobenius norm of a tensor is defined as follow:
\begin{equation}
\Vert\bm{\mathcal{X}}\Vert_F=\sqrt{\langle\bm{\mathcal{X}},\bm{\mathcal{X}}\rangle}=(\sum^{I_1}_{i_1=1}\sum^{I_2}_{i_2=1}\dots\sum^{I_N}_{i_n=1}\vert x_{i_1i_2\dots i_n}\vert ^2)^{1/2}.
\end{equation}
The mode-n tensor-matrix product of a tensor $\bm{\mathcal{X}}\in\mathbb{R}^{{I_1} \times {I_2} \times \dots\times {I_N}}$ multiplied by matrix $\textbf{U}\in\mathbb{R}^{J_n\times{I_n}}$ is denoted as $\bm{\mathcal{X}}\times{_n\textbf{U}}$:
\begin{equation}
\bm{\mathcal{X}}\times_n\textbf{U}_{i_1,i_2,\dots,i_{n-1},i_n,i_{n+1},\dots,i_N}=\sum_{i_n=1}^{I_n}{a_{{i_1}{i_2}\dots{i_N}}\times u_{i_nj_n}}
\end{equation}
where the height of matrix $\textbf{U}$ should be the same with the mode-n size of tensor $\bm{\mathcal{X}}$, this multiplication changes the mode-n size of tensor $\bm{\mathcal{X}}$ from $I_n$ to $J_n$.

We can use this kind of multiplication to change the sizes of a tensor.
\begin{equation}
\bm{\mathcal{X}}=\bm{\mathcal{A}}\times{_1}\textbf{U}^{(1)}\times{_2}\textbf{U}^{(2)}\dots\times{_N}\textbf{U}^{(N)}
\end{equation}
where the width of matrix $\textbf{U}^{(n)}$ should be equal to the size of mode-n of tensor $\bm{\mathcal{A}}$.

Through the above equations, We can introduce the decomposition of third-order tensor in this paper
\begin{equation}
\bm{\mathcal{X}}=\bm{\mathcal{G}}\times{_1}\textbf{U}^{(1)}\times{_2}\textbf{U}^{(2)}\times{_3}\textbf{U}^{(3)}
\end{equation}
where $\bm{\mathcal{X}}$ is the original tensor, and $\bm{\mathcal{G}}$ is the core tensor of $\bm{\mathcal{X}}$, $\textbf{U}^{(1)},\textbf{U}^{(2)},\textbf{U}^{(3)}$ are factor matrices where $\textbf{U}^{(n)}\textbf{U}^{(n)T}=I_{J_n},n=1,2,3$.
the core tensor can also be obtained by the reverse progress, which is given below:
\begin{equation}
\bm{\mathcal{G}}=\bm{\mathcal{X}}\times{_1}\textbf{U}^{(1)T}\times{_2}\textbf{U}^{(2)T}\times{_3}\textbf{U}^{(3)T}.
\end{equation}

\section{Correlation-based Initialization Method for TD}
Before we start our discussion about TD, we should know that TD is resolved by alternating least square (ALS) method, a third-order tensor is flattened along each modes as three matricizations, the initialized factor matrices are obtained by SVD results of matricizations. We will discuss the correlation in HSI and give a more efficient initialization method.

In this section, we first build a model of matricization and show that mode-1 and mode-2 matricizations can be well approximated a reference band. As the factor matrices $\textbf{U}^{(1)},\textbf{U}^{(2)}$ are the SVD results of mode-1 and mode-2 matricizations respectively, we resort the SVD result of reference band to approximate $\textbf{U}^{(1)},\textbf{U}^{(2)}$. Then, according our discussion, we give our correlation-based initialization method.
\subsection{Building a model of matricization}
Considering the HSI $\bm{\mathcal{X}}\in\mathbb{R}^{{I_1} \times {I_2} \times {I_3}}$ is a special tensor in which the bands are highly similar, the mode-1 matricization is actually a matrix stitched by $I_1$ slices of tensor. The Fig. 2 depicts this mode-1 matricization. The SVD-based initialization method is essentially implementing SVD on each mode-n matricization $\textbf{X}_{(n)}$ and seeking an orthonormal factor matrix $\textbf{U}^{(n)}$. The factor matrix is composed by the leading singular vectors of $\textbf{X}_{(n)}$. Since the slices in $\textbf{X}_{(1)}$ are the bands of HSI, these bands are highly correlated. The Fig. 1 shows this phenomenon. Intuitively, we can conclude that the SVD result of $\textbf{X}_{(1)}$ should be highly relevant to the SVD result of the slices in $\textbf{X}_{(1)}$. To verify our guess, we can build a model to modify the process.

We use a band which can represent the bands in HSI $\bm{\mathcal{X}}\in\mathbb{R}^{{I_1} \times {I_2} \times {I_3}}$ best as the reference band  $\overline{\textbf{X}}$. The method to achieve reference band is given in the following subsection.

The mode-1 and mode-2 matricizations $\textbf{X}_{(1)}$, $\textbf{X}_{(2)}$ can be expressed as:
\begin{equation}
\textbf{X}_{(1)}=\bm{i}_{1\times P}\otimes \overline{\textbf{X}}+\textbf{e}_1,
\end{equation}
\begin{equation}
\textbf{X}_{(2)}=\bm{j}_{1\times P}\otimes \overline{\textbf{X}}^T+\textbf{e}_2.
\end{equation}
In the equations above, $\textbf{e}$ is the approximation error, $\textbf{X}_{(1)}, \textbf{X}_{(2)}$ are approximated as the Kronecker products of a vector and the reference band $\overline{\textbf{X}}$. $\bm{i}_{1\times P}$ and $\bm{j}_{1\times P}$ are the approximation coefficient vectors. $\textbf{X}_{(3)}$ cannot be approximated by $\overline{\textbf{X}}$ for each band is viewed as a column vector in $\textbf{X}_{(3)}$. The reference band $\overline{\textbf{X}}$ can be decomposed by SVD:
\begin{equation}
\overline{\textbf{X}}=\textbf{U}\bm{\Sigma}\textbf{V}^T.
\end{equation}
Substituting $\overline{\textbf{X}}$ into Eq. 6, it is equivalent to 
\begin{equation}
 \textbf{X}_{(1)}=\bm{i}_{1\times P}\otimes (\textbf{U}\bm{\Sigma}\textbf{V}^T)+\textbf{e}_1.
\end{equation}
Since the Kronecker product has the property below:
\begin{equation}
(\textbf A\otimes \textbf B)(\textbf C\otimes \textbf D)(\textbf E\otimes \textbf F)=(\textbf{ACE})\otimes(\textbf{BDF}).
\end{equation}
The vector $\bm{i}_{1\times P}$ can be represented as:
\begin{equation}
\bm{i}_{1\times P}=1\times \bm{\alpha}^T \times \textbf B
\end{equation}
where $\bm{\alpha}^T$ is a vector, its first entry is 1, and other entries are 0. The first row of matrix $\textbf B$ is $\bm{i}_{1\times P}$. Then, we can derive:
\begin{equation}
 \textbf{X}_{(1)}=\textbf{U}(\bm{\alpha}^T\otimes \bm{\Sigma})(\textbf B\otimes \textbf{V}^T)+\textbf{e}_1.
\end{equation}
For $\textbf{X}_{(2)}$, we can derive the similar result, where $\overline{\textbf{X}}$ is transposed.
\begin{equation}
 \textbf{X}_{(2)}=\textbf{V}(\bm{c}\otimes \bm{\Sigma})(\textbf D^T\otimes \textbf{U}^T)+\textbf{e}_2.
\end{equation}

Therefore, the SVD on $\textbf{X}_{(1)}$ results in the equation above. In other words, the SVD-based initialization method obtains the same left singular matrix as $\overline{\textbf{X}}$. We can obtain the same result on $\textbf{X}_{(2)}$, because $\textbf{X}_{(2)}$ and $\textbf{X}_{(1)}$ are both approximated by $\overline{\textbf{X}}$. These two matricizations are similar because they are the bands spliced in different directions. Thus, the SVD-based initialization method is doing SVD on the duplications of $\overline{\textbf{X}}$. The $\textbf{X}_{(3)}$ cannot be handled in this way, because the band is viewed as vector and spatial correlation is not considered in this matricization, thus the SVD on the reference band is not correlated with the SVD on $\textbf{X}_{(3)}$. Our method to obtain the factor matrix $\textbf{U}^{(3)}$ is the same with the SVD-based initialized method.

\subsection{Correlation-based initialization method}
Our method is inspired by the high correlation of bands in HSIs. According to the model we build, when $\textbf{X}_{(1)}$ can be well represented by $\overline{\textbf{X}}$, the SVD results on $\textbf{X}_{(1)}$ and $\overline{\textbf{X}}$ are also relevant. Since the spectral correlation of HSIs is strong, we can conclude that $\textbf{X}_{(1)}$ can be well represented by $\overline{\textbf{X}}$. We present our method in this section.

The reference band $\overline{\textbf{X}}$ is critical important in this method, the $\overline{\textbf{X}}$ should well represent the bands in $\bm{\mathcal{X}}\in\mathbb{R}^{{I_1} \times {I_2} \times {I_3}}$. To obtain such a reference band, the representative ability and the computational cost are both considered as the rules of our method. Thus, we recommend adopt the mean band as the reference band.
\begin{algorithm}[htb]
\caption{Correlation-based TD initialization method}
\KwIn{Tensor $\bm{\mathcal{X}}\in\mathbb{R}^{{I_1} \times {I_2} \times {I_3}}$ and the size of core tensor $R_1,R_2, R_3$}.
Obtain $\overline{\textbf{X}}$ by averaging $I_3$ bands;\\
Apply SVD on $\overline{\textbf{X}}$ to obtain factor matrices $\textbf{U}^{(1)}, \textbf{U}^{(2)}$;\\
Obtain $\bm{\mathcal{X}}_3$ by Eq. 14;\\
Apply SVD on the mode-3 matricization of $\bm{\mathcal{X}}_3$ to obtain $\textbf{U}^{(3)}$;\\
\KwOut{The initialization factor matrices $\textbf{U}^{(1)}, \textbf{U}^{(2)},\textbf{U}^{(3)}$ of TD.}
\end{algorithm}
\begin{equation}
\bm{\mathcal{X}}_3=\bm{\mathcal{X}}\times_1\textbf{U}^{(1)}\times_2\textbf{U}^{(2)}.
\end{equation}
\subsection{Computational cost}
The computational cost of SVD is related to the matrix size. There are $MN^2$ times multiplication operations for applying SVD on a $M\times N$ matrix, thus, we need $I_1(I_2I_3)^2$ multiplication operations for applying SVD on the mode-1 matricization of tensor $\bm{\mathcal{X}}\in\mathbb{R}^{{I_1} \times {I_2} \times {I_3}}$. But the multiplication operations are much less in our initialization method for the size of reference band is $I_1\times I_2$,  which is only the size of one slice in mode-1 matricization. Moreover, we can obtain initial factor matrices $\textbf{U}^{(1)}, \textbf{U}^{(2)}$ by applying SVD once, especially the computational cost of applying SVD on a mode-n matricization is large. It is obvious that we can save computational cost by averaging the bands to obtain an averaged band, the final computational cost of our method is $I_1I^2_2+I_3(I_1I_2)^2$, where $I_3(I_1I_2)^2$ is the computational cost of $\textbf{U}^{(3)}$, while the SVD-based method takes $I_3(I_2I_1)^2+I_1(I_2I_3)^2+I_2(I_3I_1)^2$ times multiplication. Considering there are usually hundreds of bands in HSIs, which means $I_3$ is large, thus, the term of $I_1I_2^2$ is negligible compared to other terms. In other words, the computational cost of our method is concentrated on the solving of factor matrix $\textbf{U}^{(3)}$. The results in Table 3 and Table 4 verify our conclusion.

\section{Experimental Results}
To verify the compression performance and time consumption of our proposed correlation-based TD initialization method, the experiments were carried out using MATLAB 2015b along with Tensor Toolbox \cite{bader2015matlab} on a machine of Intel Core i7 6700 processor and 16GB RAM. We show our compression performance, initialization results and time consumption in this section.
\subsection{Datasets and evaluation metrics}
We consider two HSIs in our experiments: Indian Pines and Salinas, which are obtained by Airborne Visible/Infrared Imaging Spectrometer (AVIRIS) sensor. The sizes are $145\times145\times200$, $512\times217\times224$, respectively the first two dimensionalities are spatial sizes and the third dimensionality is the spectral size. These HSIs are obtained on a spectral range of 400-2500 nm.

The fitness of Tucker decomposition is defined as the Frobenius norm of the original tensor and reconstructed tensor. It is given as follow:
\begin{equation}
Fitness=1-\Vert\bm{\mathcal{X}}-\bm{\mathcal{X}}_{rec}\Vert_F/\Vert\bm{\mathcal{X}}\Vert_F.
\end{equation}
We compare the initialization time, the number of iterations in TD and the total time consumption of the TD which includes the initialization time and iteration time.
\subsection{Results}
We compare our method with three compression method: principal component analysis (PCA)+JPEG2000, three Dimensional-Set Partitioned Embedded bloCK (3D-SPECK) and multilinear principal component analysis (MPCA). PCA+JPEG2000 is a traditional compression method applied on HSI, 3D-SPECK is a commonly used method based on wavelet, MPCA is multilinear algebra method that can compress spectral and spatial redundancy simultaneously. These three methods are representative in three areas. Our method benefits from the TD, can well compress the HSI. The results are list in Table 2. The results show that our proposed method achieves the best compression performance.

\begin{table}
\centering
\caption{The SNR (dB) of our proposed method is compared with three methods with BR varying from 0.1 bpppb to 0.4 bpppb.}
{\begin{tabular}{lccccc} 
\toprule
&&PCA+JPEG2000&3D-SPECK&MPCA&Our method\\
\midrule
\multirow{4}{*}{Indian Pines}&0.1&\textbf{34.21}&27.02&33.58&34.16\\
&0.2&36.54&30.53&36.72&\textbf{36.89}\\
&0.3&37.51&32.71&37.98&\textbf{38.24}\\
&0.4&39.12&35.64&38.51&\textbf{40.33}\\
\midrule
\multirow{4}{*}{Salinas}&0.1&\textbf{34.87}&29.12&34.21&34.36\\
&0.2&36.65&31.57&36.67&\textbf{37.71}\\
&0.3&38.12&33.54&37.55&\textbf{39.64}\\
&0.4&39.41&34.84&38.21&\textbf{41.15}\\
 \bottomrule
\end{tabular}}
\label{sample-table}
\end{table}

To verify the initialization performance of our method, we test our method against random initialization method and SVD-based initialization method. The sizes of core tensor are all setting as the half of each corresponding dimensionalities. The results in Table 3 and Table 4 show that our initialization method consumes less time than random and SVD-based initialization methods, while the final compression performances are same (which is listed in Table 2). 
\begin{table}
\centering
\caption{The comparative results of three initialization methods on Indian Pines.}
{\begin{tabular}{lccccc} 
\toprule
&Random method&SVD-based method&Our method\\
\midrule
Initialization time (s)&-&1.502&\textbf{0.270}\\
Number of iterations&4&2&2\\
Convergence time (s)&2.940&\textbf{1.501}&1.750\\
Total time (s)&2.940&3.003&\textbf{2.020}\\
\bottomrule
\end{tabular}}
\label{sample-table}
\end{table}

\begin{table}
\centering
\caption{The comparative results of three initialization methods on Salinas.}
{\begin{tabular}{lccccc} 
\toprule
&Random method&SVD-based method&Our method\\
\midrule
Initialization time (s)&-&32.175&\textbf{3.066}\\
Number of iterations&4&2&2\\
Convergence time (s)&40.427&\textbf{30.306}&33.373\\
Total time (s)&40.427&62.481&\textbf{36.440}\\
\bottomrule
\end{tabular}}
\label{sample-table}
\end{table}

To further evaluate the initialization performance of our method, we compare our method with SVD-based method on Indian Pines. The time to obtain each factor matrix is listed in Table 5. Our method obtains the first two factor matrices $\textbf{U}^{(1)}$, $\textbf{U}^{(2)}$ in 0.625 seconds while the SVD-based method needs 199.376 seconds in total. Thus, this time consumption of 0.625 second can be ignored and most of time is consumed for computing the $\textbf{U}^{(3)}$ in our method, which is still much less than the SVD-based method. Although, we carry out SVD on a mode-3 matricization, the matricization is obtained by the original tensor rotated by the first two factor matrices and the sizes of tensor become smaller, so that the SVD consumes less time in our method.
\begin{table}
\centering
\caption{The time (s) consumption of computing three factor matrices in 100 times.}
{\begin{tabular}{lcc} 
\toprule
Factor matrix&SVD-based method&Our method\\
\midrule
$\textbf{U}^{(1)}$&93.438&\multirow{2}{*}{\textbf{0.625}}\\
$\textbf{U}^{(2)}$&105.938&\\
$\textbf{U}^{(3)}$&110.780&\textbf{41.125}\\
Total cost&310.156&\textbf{41.750}\\
\bottomrule
\end{tabular}}
\label{sample-table}
\end{table}
\subsection{Iterations and fitness}
We also compare the fitness of three method varying with iterations. As the fitness of random method is much lower than our method and SVD-based method, we show the curve from the second iteration. Fig. 3 shows that our method has a fitness close to SVD-based method and much better than random method. Our method saves nearly three iterations compared to the random method. We compare the fitness difference of each iteration and the curve in logarithmic coordinate system shows that our method has a performance close to SVD-based method.
\begin{figure}
\centering
\subfigure{%
\resizebox*{10cm}{!}{\includegraphics{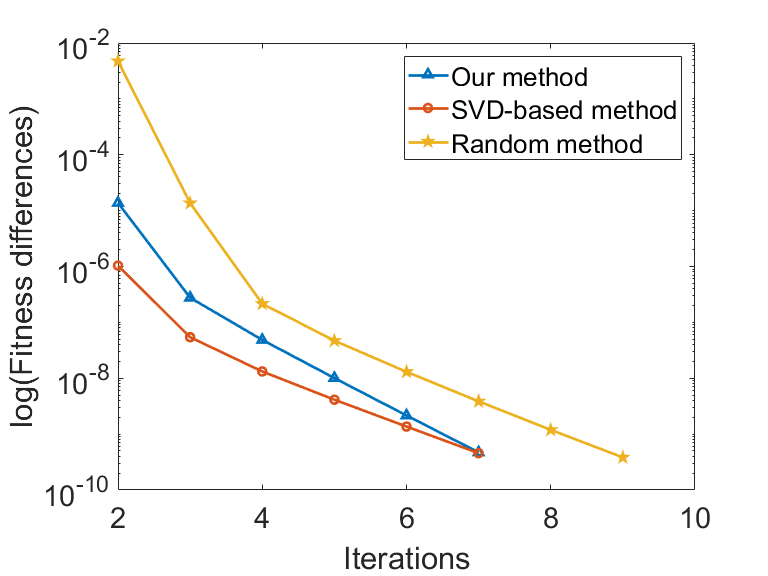}}}\hspace{5pt}
\caption{The fitness difference varies with the iterations.} \label{sample-figure}
\end{figure}

We also list the fitness differences of first three iterations in Table 6 for comparison.
\begin{table}
\centering
\caption{The fitness differences of three methods in first three iterations.}
{\begin{tabular}{lccccc} 
\toprule
Iteration&SVD-based method&Random method&Our method\\
\midrule
1&1.034e-6&4.712e-3&1.374e-5\\
2&5.423e-8&1.363e-5&2.761e-7\\
3&1.321e-8&2.158e-7&4.855e-8\\
\bottomrule
\end{tabular}}
\label{sample-table}
\end{table}	
The results show the fitness changing of our method is close to the SVD-based method and much smaller than Random method. This verifies that our method can obtain the initialized factor matrices close to the best initialized factor matrices with a much lower computational cost.
\section{Conclusion}
We utilize the strong correlation between the bands of HSIs and propose a new initialization method for TD, we prove that mode-1 and mode-2 matricizations of HSI can be well approximated by the Kronecker product of a reference band and an approximation vector. According to this conclusion, we propose our correlation-based initialization algorithm for tensor-based HSI compression. Our method ensures the TD method can converge in a few iterations and the computational cost is obviously saved. The experiments verify the good initialization performance, compression performance and less time consumption of our method.
\section*{Acknowledgement(s)}
This work is supported in part by the Open Research Fund of Key Laboratory of Spectral Imaging Technology, Chinese Academy of Sciences (Grant No. LSIT201606D) and the Open Project Program of the National Laboratory of Pattern Recognition (NLPR) (Grant No. 201800030).

\bibliographystyle{spbasic}
\bibliography{interactcadsample}

\begin{thebibliography}{19}
\providecommand{\natexlab}[1]{#1}
\providecommand{\url}[1]{{#1}}
\providecommand{\urlprefix}{URL }
\expandafter\ifx\csname urlstyle\endcsname\relax
  \providecommand{\doi}[1]{DOI~\discretionary{}{}{}#1}\else
  \providecommand{\doi}{DOI~\discretionary{}{}{}\begingroup
  \urlstyle{rm}\Url}\fi
\providecommand{\eprint}[2][]{\url{#2}}

\bibitem[{Akbari et~al.(2010)Akbari, Kosugi, Kojima, and
  Tanaka}]{akbari2010detection}
Akbari H, Kosugi Y, Kojima K, Tanaka N (2010) Detection and analysis of the
  intestinal ischemia using visible and invisible hyperspectral imaging. IEEE
  Transactions on Biomedical Engineering 57(8):2011--2017

\bibitem[{Bader et~al.(2015)Bader, Kolda et~al.}]{bader2015matlab}
Bader BW, Kolda TG, et~al. (2015) Matlab tensor toolbox version 2.6. available
  online. URL: http://www sandia gov/tgkolda/TensorToolbox

\bibitem[{Cichocki et~al.(2008)Cichocki, Zdunek, and
  Amari}]{cichocki2008nonnegative}
Cichocki A, Zdunek R, Amari Si (2008) Nonnegative matrix and tensor
  factorization [lecture notes]. IEEE Signal Processing Magazine 25(1):142--145

\bibitem[{Coutinho et~al.(2017)Coutinho, Cintra, and Bayer}]{coutinho2017low}
Coutinho VdA, Cintra RJ, Bayer FM (2017) Low-complexity multidimensional dct
  approximations for high-order tensor data decorrelation. IEEE Transactions on
  Image Processing 26(5):2296--2310

\bibitem[{Da~Silva~Fernandes et~al.(2017)Da~Silva~Fernandes, Tork, and
  da~Gama}]{da2017initialization}
Da~Silva~Fernandes S, Tork HF, da~Gama JMP (2017) The initialization and
  parameter setting problem in tensor decomposition-based link prediction. In:
  2017 IEEE International Conference on Data Science and Advanced Analytics
  (DSAA), IEEE, pp 99--108

\bibitem[{Dragotti et~al.(2000)Dragotti, Poggi, and
  Ragozini}]{dragotti2000compression}
Dragotti PL, Poggi G, Ragozini AR (2000) Compression of multispectral images by
  three-dimensional spiht algorithm. IEEE Transactions on Geoscience and Remote
  Sensing 38(1):416--428

\bibitem[{Fang et~al.(2017)Fang, Wang, Li, and
  Benediktsson}]{fang2017hyperspectral}
Fang L, Wang C, Li S, Benediktsson JA (2017) Hyperspectral image classification
  via multiple-feature-based adaptive sparse representation. IEEE Transactions
  on Instrumentation and Measurement 66(7):1646--1657

\bibitem[{Jiang et~al.(2018)Jiang, Ding, Tang, and Luo}]{jiang2018image}
Jiang B, Ding C, Tang J, Luo B (2018) Image representation and learning with
  graph-laplacian tucker tensor decomposition. IEEE Transactions on Cybernetics
  49(4):1417--1426

\bibitem[{Karami et~al.(2012)Karami, Yazdi, and
  Mercier}]{karami2012compression}
Karami A, Yazdi M, Mercier G (2012) Compression of hyperspectral images using
  discerete wavelet transform and tucker decomposition. IEEE Journal of
  Selected Topics in Applied Earth Observations and Remote Sensing
  5(2):444--450

\bibitem[{Kolda and Bader(2009)}]{kolda2009tensor}
Kolda TG, Bader BW (2009) Tensor decompositions and applications. SIAM Review
  51(3):455--500

\bibitem[{Ma et~al.(2018)Ma, Li, Li, Mei, and Ma}]{ma2018hyperspectral}
Ma Y, Li C, Li H, Mei X, Ma J (2018) Hyperspectral image classification with
  discriminative kernel collaborative representation and tikhonov
  regularization. IEEE Geoscience and Remote Sensing Letters 15(4):587--591

\bibitem[{Md~Noor et~al.(2017)Md~Noor, Ren, Marshall, and
  Michael}]{md2017hyperspectral}
Md~Noor S, Ren J, Marshall S, Michael K (2017) Hyperspectral image enhancement
  and mixture deep-learning classification of corneal epithelium injuries.
  Sensors 17(11):2644

\bibitem[{Mei et~al.(2018)Mei, Ma, Li, Fan, Huang, and Ma}]{mei2018robust}
Mei X, Ma Y, Li C, Fan F, Huang J, Ma J (2018) Robust gbm hyperspectral image
  unmixing with superpixel segmentation based low rank and sparse
  representation. Neurocomputing 275:2783--2797

\bibitem[{Shi and Pun(2017)}]{shi20173d}
Shi C, Pun CM (2017) 3d multi-resolution wavelet convolutional neural networks
  for hyperspectral image classification. Information Sciences 420:49--65

\bibitem[{Sidiropoulos et~al.(2017)Sidiropoulos, De~Lathauwer, Fu, Huang,
  Papalexakis, and Faloutsos}]{sidiropoulos2017tensor}
Sidiropoulos ND, De~Lathauwer L, Fu X, Huang K, Papalexakis EE, Faloutsos C
  (2017) Tensor decomposition for signal processing and machine learning. IEEE
  Transactions on Signal Processing 65(13):3551--3582

\bibitem[{Tucker(1966)}]{tucker1966some}
Tucker LR (1966) Some mathematical notes on three-mode factor analysis.
  Psychometrika 31(3):279--311

\bibitem[{Wang et~al.(2017)Wang, Lin, Zhao, Yue, Meng, and
  Leung}]{wang2017compressive}
Wang Y, Lin L, Zhao Q, Yue T, Meng D, Leung Y (2017) Compressive sensing of
  hyperspectral images via joint tensor tucker decomposition and weighted total
  variation regularization. IEEE Geoscience and Remote Sensing Letters
  14(12):2457--2461

\bibitem[{Zeng et~al.(2017)Zeng, Zhang, and Bai}]{zeng2017method}
Zeng W, Zhang X, Bai Y (2017) Method for multispectral images denoising based
  on tensor-singular value decomposition. Journal of Applied Remote Sensing
  11(3):035019

\bibitem[{Zhang et~al.(2016)Zhang, Zhang, Jiao, Liu, Wang, and
  Hou}]{zhang2016weighted}
Zhang E, Zhang X, Jiao L, Liu H, Wang S, Hou B (2016) Weighted multifeature
  hyperspectral image classification via kernel joint sparse representation.
  Neurocomputing 178:71--86

\end{thebibliography}
%
%
%
%

\ifCLASSOPTIONcaptionsoff

\fi


%




\begin{IEEEbiographynophoto}{Rui Li}
received the B. S. and M. S. degree from Xidian University, Xi'an, P. R. China, in 2012 and 2015 respectively. 
He is currently working towards the Ph. D. degree in School of Electronic and Information Engineering at Xi'an Jiaotong University, Xi'an, P. R. China. His research interests include vector quantization and hyper-spectral image processing.
\end{IEEEbiographynophoto}
\begin{IEEEbiographynophoto}{Zhibin Pan}
received the B. S. degree in Information and Telecommunication Engineering in 1985 and the M. S. degree in Automation Science and Technology in 1988 from Xi'an Jiaotong University, P. R. China, respectively. He received the Ph.  D. degree in Electrical Engineering in 2000 from Tohoku University, Japan. He is a professor in the Department of Information and Telecommunication Engineering, Xi'an Jiaotong University, P. R. China. His current research interests include image compression, multimedia security and object recognition.
\end{IEEEbiographynophoto}
\begin{IEEEbiographynophoto}{Yang Wang}
received his B. S. degree from Xi'an Jiaotong University, Xi'an, P. R. China, in 2010.
He is currently working towards the Ph. D. degree in School of Electronic and Information Engineering at Xi'an Jiaotong University, Xi'an, P. R. China. His research interests include image coding and image processing.
\end{IEEEbiographynophoto}





\end{document}